\documentclass{PoS}

\title{Wgamma and Zgamma production at the LHC in NNLO QCD}

\ShortTitle{Wgamma and Zgamma production at the LHC in NNLO QCD}

\author{Massimiliano Grazzini\\
        University of Zurich\\
        E-mail: \email{grazzini@physik.uzh.ch}}

\author{\speaker{Stefan Kallweit}\\%
        Johannes-Gutenberg-University Mainz\\
        E-mail: \email{kallweit@uni-mainz.de}}

\author{Dirk Rathlev\\
        DESY Hamburg\\
        E-mail: \email{dirk.rathlev@desy.de}}

\abstract{
We consider the production of $W\gamma$ and $Z\gamma$ pairs at the LHC,
and report on the fully differential computation of next-to-next-to-leading order (NNLO) corrections in QCD perturbation theory.
The calculation includes leptonic vector-boson decays with the corresponding spin correlations,
off-shell effects and final-state photon radiation.
We present numerical results for $pp$ collisions at $7\;\TeV$, and compare them with available ATLAS data.
In the case of $Z\gamma$ production, the impact of NNLO corrections is generally moderate, ranging from $8\%$ to $17\%$, depending on the applied cuts. In the case of $W\gamma$ production, the NNLO effects are more important, and range from $19\%$ to $26\%$, thereby improving the agreement of the theoretical predictions with the data.
As expected, a veto against jets significantly reduces the impact of QCD radiative corrections.
}

\FullConference{12th International Symposium on Radiative Corrections (Radcor 2015) and LoopFest XIV (Radiative Corrections for the LHC and Future Colliders)\\
		 15-19 June  2015\\
		 UCLA Department of Physics \& Astronomy Los Angeles, CA, USA}

\usepackage{amsmath,url}
\usepackage{graphicx}
\usepackage{multirow}
\usepackage{color}
\usepackage{amssymb}
\usepackage{xspace}

\newcommand\as{\alpha_{\mathrm{S}}}

\def\to{\rightarrow}

\def\pT{p_\mathrm{T}}
\def\mT{m_\mathrm{T}}

\newcommand{\GeV}{\mbox{GeV}}
\newcommand{\TeV}{\mbox{TeV}}

\newcommand{\qqbar}{q\bar{q}}
\newcommand{\qqbarprime}{q{\bar q^\prime}}

\def\reffi#1{\mbox{Figure~\ref{#1}}}

\def\refta#1{\mbox{Table~\ref{#1}}}

\def\citere#1{\mbox{Ref.~\cite{#1}}}
\def\citeres#1{\mbox{Refs.~\cite{#1}}}

\newcommand{\qT}{q_{\mathrm{T}}}

\newcommand{\Matrix}{{\rmfamily \scshape Matrix}\xspace}
\newcommand{\Munich}{{\rmfamily \scshape Munich}\xspace}

\begin{document}

\setcounter{footnote}{1}
\renewcommand{\thefootnote}{\fnsymbol{footnote}}

\section{Introduction}

The discovery of a new scalar resonance in the search for the Standard Model~(SM) Higgs 
boson~\cite{Aad:2012tfa,Chatrchyan:2012ufa}
is a milestone in the LHC physics programme. The properties of this new particle
closely resemble those of the Higgs boson, but further work is needed to clarify if it is really the 
Higgs boson predicted by the SM, or something (slightly) different.
Vector-boson pair production has a prominent role in this context. It represents an irreducible background
to Higgs and new-physics searches, and, at the same time, it provides information on the form and the strength
of the vector-boson gauge couplings.
The interactions of $W$ and $Z$ bosons with photons
are particularly interesting as they test the
$WW\gamma$ and $ZZ\gamma$ couplings, which are predicted by the non-Abelian $SU(2)_L\otimes U(1)_Y$ gauge group.

The high-energy proton--proton collisions at the LHC allow us to
explore the production of $V\gamma$ ($V=W^\pm,Z$) pairs in a new energy domain. Measurements of 
$V\gamma$ final states have been carried out by ATLAS~\cite{Aad:2011tc,Aad:2012mr,Aad:2013izg,Aad:2014fha} 
and CMS~\cite{Chatrchyan:2011rr,Chatrchyan:2013fya,Chatrchyan:2013nda,Khachatryan:2015kea} using the data sets 
at centre-of-mass energy $\sqrt{s}=7\;\TeV$ and $8\;\TeV$. These measurements have been compared to the SM predictions and used to improve
the limits on anomalous couplings and on the production of possible new resonances.

The most precise SM predictions in fixed-order perturbation theory available for $V\gamma$ production at hadron colliders are, on the one hand side, electroweak corrections at next-to-leading order (NLO), which were presented in \citere{Denner:2014bna} for $W\gamma$ and in \citere{Denner:2015fca} for $Z\gamma$ production, and, on the other side, next-to-next-to-leading order (NNLO) QCD corrections, which were discussed in \citere{Grazzini:2015nwa}.\footnote{First results for $Z\gamma$ production were presented in \citere{Grazzini:2013bna}, and for $W\gamma$ production in \citere{Grazzini:2014pqa}.}
Full leptonic decays, off-shell effects and final-state photon radiation are consistently included in all of these calculations, i.e.\ the hadronic production of the full final states $\ell^+\ell^-\gamma$ and $\nu\bar\nu\gamma$ as well as $\nu \ell^+\gamma$ and $\ell^-\bar\nu\gamma$ is evaluated, and often referred to as $Z\gamma$ and $W\gamma$ production for convenience.

In these proceedings we discuss selected results on $Z\gamma$ and $W\gamma$ fiducial cross sections and distributions at NNLO QCD accuracy, and provide comparisons to ATLAS $\sqrt{s}=7\;\TeV$ data. All results shown here were presented in \citere{Grazzini:2015nwa}.

\section{Details of the calculation}
\label{sec:outline}

The NNLO computation requires the evaluation of tree-level scattering amplitudes with up to two additional (unresolved) partons, 
of one-loop amplitudes with up to one additional (unresolved) parton, 
and of one-loop squared and two-loop corrections to the Born subprocess
($\qqbar\to \ell^+\ell^-\gamma$ and $\qqbar\to\nu_\ell{\bar \nu_\ell}\gamma$ for $Z\gamma$, 
$\qqbarprime\to \ell\nu_\ell\gamma$ for $W\gamma$).
Furthermore, processes with charge-neutral final states receive loop-induced contributions from the gluon-fusion channel 
($gg\to \ell^+\ell^-\gamma$ and $gg\to\nu_\ell{\bar \nu_\ell}\gamma$).
In our computation, all required tree-level and one-loop amplitudes are obtained from
the {\sc OpenLoops} generator~\cite{Cascioli:2011va}\footnote{The {\sc OpenLoops} one-loop generator
by F.~Cascioli, J.~Lindert, P.~Maierh{\"o}fer and S.~Pozzorini
is publicly available at \url{http://openloops.hepforge.org}.}, which
implements
a fast numerical recursion 
for the calculation of NLO scattering amplitudes within the SM. For the numerically stable evaluation of tensor integrals we rely on the {\sc Collier} library~\cite{Denner:2014gla}, which is based on the Denner--Dittmaier reduction techniques~\cite{Denner:2002ii,Denner:2005nn} and the scalar integrals of~\cite{Denner:2010tr}.

The two-loop corrections to the Drell--Yan-like Born processes, where the photon
is radiated off the final-state leptons, have been available for a long time~\cite{Matsuura:1988sm}.
The last missing ingredient,
the genuine two-loop corrections to the $V\gamma$ amplitudes,
have been presented in \citere{Gehrmann:2011ab}.

The bookkeeping of all partonic subprocesses and the numerical integration of the different cross section contributions is managed by the fully automatized 
\Munich framework\footnote{\Munich is the abbreviation of ``MUlti-chaNnel Integrator at 
Swiss~(CH) precision''---an automated parton level NLO generator by S.~Kallweit. In preparation.}, which also automatically organizes the mediation of NLO-like soft and collinear divergences by means of the Catani--Seymour dipole subtraction method~\cite{Catani:1996jh,Catani:1996vz}.
To deal with NNLO corrections to the hadronic production of arbitrary colourless final states $F$, the $\qT$ subtraction formalism~\cite{Catani:2007vq} has been implemented into this framework, i.e.\ the extraction procedures for all required counterterms and hard-collinear coefficients up to $\mathcal{O}(\alpha_{\mathrm{s}}^2)$, which were presented in \citeres{Bozzi:2005wk,deFlorian:2001zd,Catani:2011kr,Catani:2012qa,Catani:2013tia}, were added in a process-independent way, giving rise to the numerical program \Matrix\footnote{\Matrix{} is the abbreviation of 
``\Munich{} Automates qT subtraction and Resummation
to Integrate Cross Sections'', by M.~Grazzini, S.~Kallweit, D.~Rathlev, M.~Wiesemann. 
In preparation.}. This tool has been applied to several hadronic processes at inclusive and fully differential level~\cite{Grazzini:2015nwa,Grazzini:2013bna,Cascioli:2014yka,Gehrmann:2014fva,Grazzini:2015hta}, and also in the first combination of next-to-next-to-leading logarithmic resummation with NNLO fixed-order accuracy for on-shell WW and ZZ production~\cite{Grazzini:2015wpa}. More calculational details are described in \citere{Grazzini:2015nwa}.

\section{Numerical results}
\label{sec:results}
\label{setup}
For the electroweak couplings we use the so-called $G_\mu$ scheme,
where the input parameters are $G_F$, $m_W$, $m_Z$. In particular we 
use the values
$G_F = 1.16639\times 10^{-5}\;\GeV^{-2}$, $m_W=80.399\;\GeV$,
$m_Z = 91.1876\;\GeV$, $\Gamma_Z=2.4952\;\GeV$ and $\Gamma_W=2.1054\;\GeV$. We set the CKM matrix to unity.
We use the MMHT 2014~\cite{Harland-Lang:2014zoa} sets of parton distribution functions (PDFs), with
densities and $\as$ evaluated at each corresponding order
(i.e., we use $(n+1)$-loop $\as$ at N$^n$LO, with $n=0,1,2$),
and we consider $N_f=5$ massless quarks/antiquarks and gluons in 
the initial state.
The default renormalization ($\mu_R$) and factorization ($\mu_F$) scales are set to
$\mu_R=\mu_F=\mu_0\equiv\sqrt{m_V^2+(\pT^{\gamma})^2}$, and scale uncertainties are estimated by varying $\mu_F$ and $\mu_R$ independently
in the range $0.5\mu_0$ and $2\mu_0$.

The present formulation of the $\qT$ subtraction formalism~\cite{Catani:2007vq}
is limited to the production of colourless systems $F$ and, hence, it does not
allow us to deal with the 
parton fragmentation subprocesses.
Therefore, we consider only direct photons, and 
we rely on the smooth cone isolation criterion~\cite{Frixione:1998jh}. Considering a cone 
of radius $r=\sqrt{(\Delta \eta)^2+(\Delta \phi)^2}$ around the photon, 
we require that the total amount of hadronic (partonic) transverse energy $E_T$ 
inside the cone is smaller than $E_{T}^{\rm max}(r)$,
\begin{equation}
E_{T}^{\rm max}(r) \equiv  \epsilon_\gamma \,\pT^\gamma \left(\frac{1-\cos r}{1- \cos R}\right)^n \, ,
\label{eq:frixione}
\end{equation}
where $\pT^\gamma$ is the photon transverse momentum; the isolation criterion
$E_T < E_{T}^{\rm max}(r)$ has to be fulfilled for all cones with $r\leq R$.
All results presented here are obtained with $\epsilon_\gamma=0.5$, $n=1$ and $R=0.4$.
For these results, we verified at NLO that the difference between using 
smooth and hard cone isolation is at the $1-2\%$ level\footnote{Obviously, the agreement also significantly depends on the fragmentation 
function used when employing the hard cone isolation, which typically has large uncertainties.},
i.e.\ well below the current experimental uncertainties and still smaller than
the remaining theoretical uncertainties.
We can thus safely compare our theoretical predictions with experimental data.

Jets are reconstructed with the anti-$k_T$ algorithm~\cite{Cacciari:2008gp} with radius parameter $D=0.4$, and a jet must have $\pT^{\rm jet}>30\;\GeV$ and $|\eta^{\rm jet}|<4.4$. 

In these proceedings, we limit ourselves to compare our predictions to the ATLAS results for $W\gamma$ and $Z\gamma$ at $7\;\TeV$~\cite{Aad:2013izg}.
Experimental results and theoretical predictions on fiducial cross sections are collected in \refta{tab:results_VA} for the different channels, with and without a veto against jets.
The precise kinematic cuts to define these fiducial cross sections are detailed in Tables 1,4,6 of \citere{Grazzini:2015nwa}, and are not repeated here.

\begin{table}[t]
\begin{center}
\setlength{\tabcolsep}{4pt}
\begin{tabular}{|c r r r r r r r r|}
\hline
\multicolumn{9}{|c|}{}\\[-2.5ex]
\multicolumn{1}{|c}{$\begin{array}{c}p^\gamma_{\rm{T,cut}}\\\mathrm{[GeV]}\end{array}$}
& $N_{\rm{jet}}$ &
& \multicolumn{1}{c}{$\begin{array}{c}\sigma_{\mathrm{LO}}\\\mathrm{[pb]}\end{array}$}
& \multicolumn{1}{c}{$\begin{array}{c}\sigma_{\mathrm{NLO}}\\\mathrm{[pb]}\end{array}$}
& \multicolumn{1}{c}{$\begin{array}{c}\sigma_{\mathrm{NNLO}}\\\mathrm{[pb]}\end{array}$}
& \multicolumn{1}{c}{$\begin{array}{c}\sigma_{\mathrm{ATLAS}}\\\mathrm{[pb]}\end{array}$}
& $\dfrac{\sigma_{\mathrm{NLO}}}{\sigma_{\mathrm{LO}}}$
& $\dfrac{\sigma_{\mathrm{NNLO}}}{\sigma_{\mathrm{NLO}}}$
\\[2ex]

\hline
\multicolumn{9}{c}{}\\[-1.5ex]
\multicolumn{9}{c}{$pp (\to W\gamma)\to\ell\nu\gamma+X\;@\;\sqrt{s}=7\,\mathrm{TeV}$}\\[.5ex]
\hline
& $\geq 0$ &
& 
& {$2.058\,^{+6.8\%}_{-6.8\%}$}
& {$2.453\,^{+4.1\%}_{-4.1\%}$}
& {$2.77\hspace*{0em}\begin{array}{l}\\[-3ex]\scriptstyle\pm0.03~{\rm(stat)}\\[-1ex]\scriptstyle\pm0.33~{\rm(syst)}\\[-1ex]\scriptstyle\pm0.14~{\rm(lumi)}\\[1ex]\end{array}\!$}
& {+136\%}%
& {\phantom{0}+19\%}%
\\[-2ex]
15 & & & $0.8726\,^{+6.8\%}_{-8.1\%}$ &&&&&\\[-2ex] 
& $= 0$ & &  
& {$1.395\,^{+5.2\%}_{-5.8\%}$}
& {$1.493\,^{+1.7\%}_{-2.7\%}$}
& {$1.76\hspace*{0em}\begin{array}{l}\scriptstyle\pm0.03~{\rm(stat)}\\[-1ex]\scriptstyle\pm0.21~{\rm(syst)}\\[-1ex]\scriptstyle\pm0.08~{\rm(lumi)}\\[1ex]\end{array}\!$}
& \phantom{0}+60\%%
& \phantom{00}+7\%%
\\[-1ex]
\hline
\multicolumn{9}{|c|}{}\\[-2ex]
40
& $\geq 0$ &
& $0.1158\,^{+2.6\%}_{-3.7\%}$ 
& {$0.3959\,^{+9.0\%}_{-7.3\%}$}
& {$0.4971\,^{+5.3\%}_{-4.7\%}$}
&
& {+242\%}%
& {\phantom{0}+26\%}%
\\[1ex]
\hline

\hline
\multicolumn{9}{c}{}\\[-1.5ex]
\multicolumn{9}{c}{$pp (\to Z\gamma)\to\ell^+\ell^-\gamma+X\;@\;\sqrt{s}=7\,\mathrm{TeV}$}\\[.5ex]
\hline
& $\geq 0$ &
& 
& {$1.222\,^{+4.2\%}_{-5.3\%}$}
& {$1.320\,^{+1.3\%}_{-2.3\%}$}
& {$1.31\hspace*{0em}\begin{array}{l}\\[-3ex]\scriptstyle\pm 0.02~{\rm(stat)}\\[-1ex]\scriptstyle\pm 0.11~{\rm(syst)}\\[-1ex] \scriptstyle\pm 0.05~{\rm(lumi)}\\[1ex]\end{array}\!$}
& {\phantom{0}+50\%}%
& {\phantom{00}+8\%}%
\\[-2ex]
15 & & & $0.8149\,^{+8.0\%}_{-9.3\%}$ &&&&&\\[-2ex] 
& $= 0$ &
&  
& {$1.031\,^{+2.7\%}_{-4.3\%}$}
& {$1.059\,^{+0.7\%}_{-1.4\%}$}
& {$1.05\hspace*{0em}\begin{array}{l}\\[-3ex]\scriptstyle\pm 0.02~{\rm(stat)}\\[-1ex]\scriptstyle\pm 0.10~{\rm(syst)}\\[-1ex]\scriptstyle\pm 0.04~{\rm(lumi)}\\[1ex]\end{array}\!$}
& \phantom{0}+27\%%
& \phantom{00}+3\%%
\\[-1ex]
\hline
\multicolumn{9}{|c|}{}\\[-2ex]
40
& $\geq 0$ &
& $0.0736\,^{+3.4\%}_{-4.5\%}$
& {$0.1320\,^{+4.2\%}_{-4.0\%}$}
& {$0.1543\,^{+3.1\%}_{-2.8\%}$}
&
& {\phantom{0}+79\%}%
& {\phantom{0}+17\%}%
\\[1ex]

\hline
\multicolumn{9}{c}{}\\[-1.5ex]
\multicolumn{9}{c}{$pp (\to Z\gamma)\to\nu\bar\nu\gamma+X\;@\;\sqrt{s}=7\,\mathrm{TeV}$}\\[.5ex]
\hline
& $\geq 0$ &
& 
& {$0.1237\,^{+4.1\%}_{-3.1\%}$}
& {$0.1380\,^{+2.5\%}_{-2.3\%}$}
& {$0.133\hspace*{0em}\begin{array}{l}\\[-3ex]\scriptstyle\pm 0.013~{\rm(stat)}\\[-1ex]\scriptstyle\pm0.020~{\rm(syst)}\\[-1ex]\scriptstyle\pm0.005~{\rm(lumi)}\\[1ex]\end{array}\!$}
& {\phantom{0}+57\%}%
& {\phantom{0}+12\%}%
\\[-2ex]
100 & & & $0.0788\,^{+0.3\%}_{-0.9\%}$ &&&&&\\[-2ex]  
\multicolumn{1}{c}{}
& $= 0$ &
&  
& {$0.0881\,^{+1.2\%}_{-1.3\%}$}
& {$0.0866\,^{+1.0\%}_{-0.9\%}$}
& {$0.116\hspace*{0em}\begin{array}{l}\\[-3ex]\scriptstyle\pm0.010~{\rm(stat)}\\[-1ex]\scriptstyle\pm0.013~{\rm(syst)}\\[-1ex]\scriptstyle\pm0.004~{\rm(lumi)}\\[1ex]\end{array}\!$}
& {\phantom{0}+12\%}%
& \multicolumn{1}{r}{\phantom{00}$-2$\%}%
\\[-7ex]
\multicolumn{9}{|c|}{}\\
\multicolumn{9}{|c|}{}\\
\hline
\end{tabular}

\end{center}
\caption{Results on fiducial cross sections to the ATLAS $7\;\TeV$ analyses on $pp\to \ell\nu\gamma$, $pp\to \ell\ell\gamma$, and $pp\to \nu\nu\gamma$. Event-selection criteria are detailed in Tables 1,4,6 of \citere{Grazzini:2015nwa}.}
\label{tab:results_VA}
\end{table}

The predicted inclusive $W\gamma$ cross sections ($W^+\gamma$ and $W^-\gamma$ are always summed over) with the soft $\pT^\gamma$ cut of $15\;\GeV$ are quite large: the NLO $K$ factor is $+136\%$, and the NNLO corrections increase the NLO results by $+19\%$.
The measurement of the inclusive cross section by ATLAS
shows a $2\sigma$ excess with respect to the NLO prediction, which is reduced to well below $1\sigma$ when including the NNLO corrections.
The impact of QCD corrections at NLO and NNLO is reduced to
$60\%$ and $7\%$, respectively, when a jet veto is applied ($N_{\rm jet}=0$).
Such an effect is expected~\cite{Catani:2001cr} and apparently leads to a more stable perturbative prediction,
but also to the possible need of more conservative procedures to estimate perturbative uncertainties.
In the exclusive case, the excess of the measured fiducial cross sections over the theoretical prediction is reduced from $1.6\sigma$ to $1.2\sigma$ when going from NLO to NNLO.
We note that the scale variations at NLO significantly underestimate the impact of the NNLO corrections, in particular in the inclusive case. 

The predicted $Z\gamma$ cross sections in the visible $Z$ decay mode with the soft $\pT^\gamma$ cut of $15\;\GeV$ 
get corrected by $+50\%$ ($+27\%$) at NLO and by $+8\%$ ($+3\%$) at NNLO
in the inclusive (exclusive) case, respectively.
Both the NLO and NNLO predictions are in agreement with the experimental results,
and the NNLO corrections improve the agreement, especially in the inclusive case.

It is obvious that the $W\gamma$ process features much larger radiative effects with respect to the $Z\gamma$ process, which should be contrasted to what happens in the case of inclusive $W$ and $Z$ boson production,
where QCD radiative corrections are essentially identical~\cite{Hamberg:1990np}. It is thus the emission of the additional
photon that breaks the similarity between the charged-current and the neutral-current processes.
By studying the LO contributions to the $Z\gamma$ and $W\gamma$ cross sections it turns out that the additional Feynman diagram in which the photon is radiated off the $W$ boson gives rise to a \textit{radiation zero}~\cite{Mikaelian:1979nr}, which does not exist in $Z\gamma$ production. This exact zero, present
in the on-shell partonic $W\gamma$ tree-level amplitude at $\cos\theta^*=1/3$, where $\theta^*$ is the scattering angle in the centre-of-mass frame, gets diluted by the convolution with the parton densities and by off-shell effects, but it is responsible for the suppression of the Born level $W\gamma$ cross section with respect to $Z\gamma$. 
 \begin{figure}[tp]
  \centering
  \includegraphics[width=0.45\textwidth]{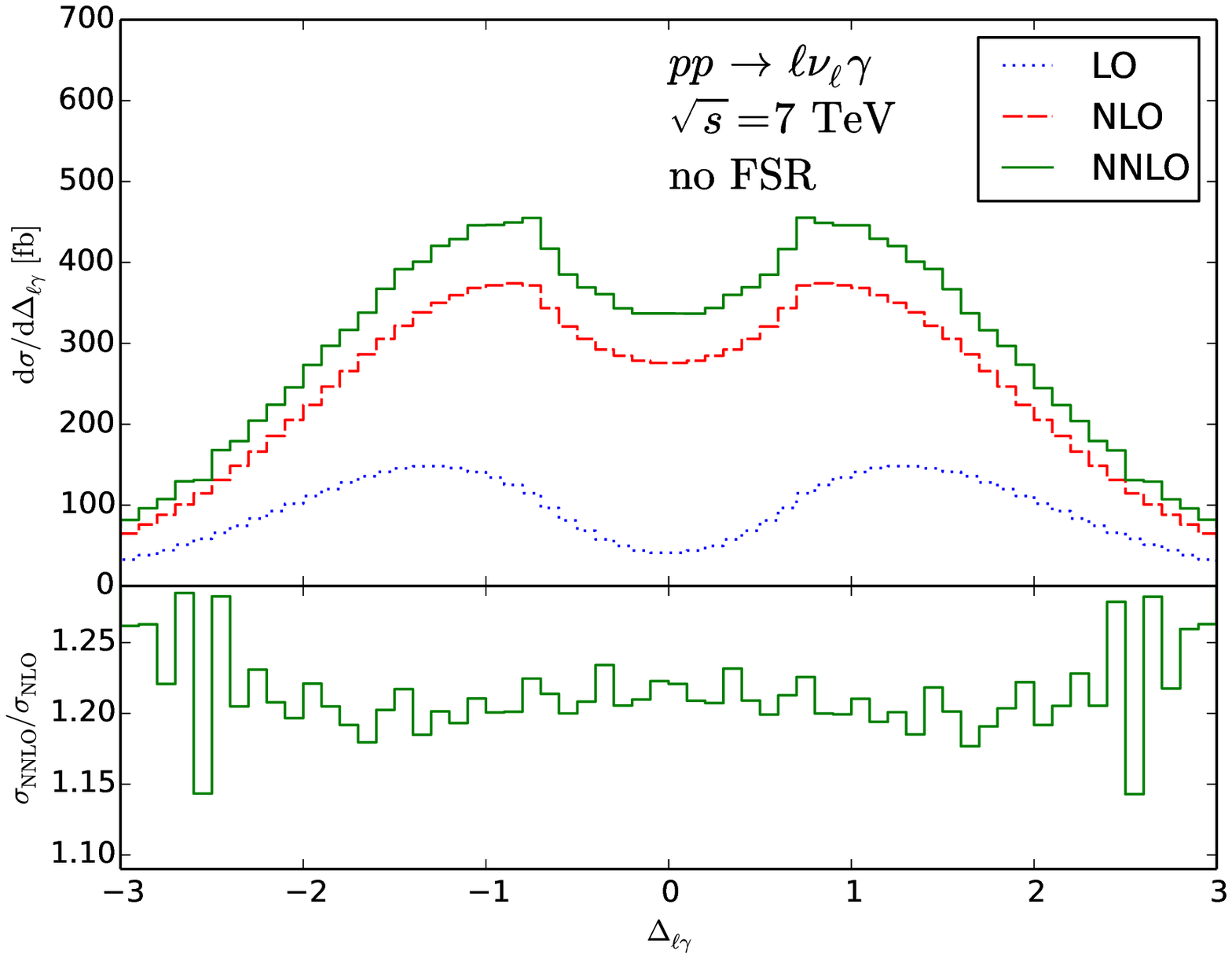}
  \hfill
  \includegraphics[width=0.45\textwidth]{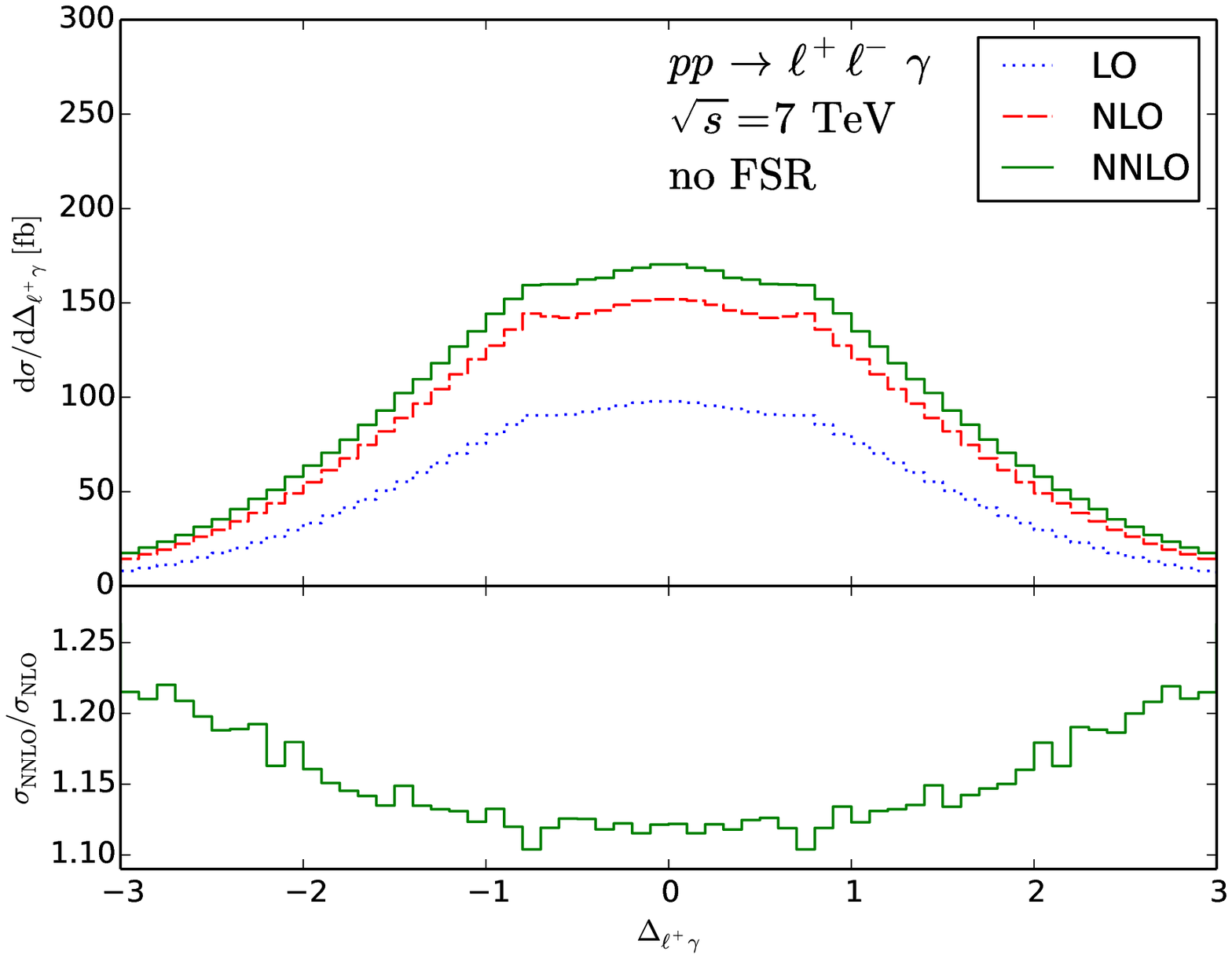}\\[-1ex]
  \caption{Rapidity difference between the charged lepton and the photon for $W\gamma$ (left) and $Z\gamma$ 
production (right) at LO (blue, dotted), NLO (red, dashed) and NNLO (green, solid). 
The lower panel shows the NNLO/NLO ratio.
Final-state radiation has been disabled for these plots.}
  \label{fig:deta}
\end{figure}
As pointed out in \citere{Baur:1994sa}, this radiation zero leads to a dip in the LO
distribution in the rapidity difference $\Delta y_{\ell\gamma}$ between the charged lepton and the photon, which is illustrated in \reffi{fig:deta}.
Real radiation appearing beyond LO breaks the radiation zero, and thus the relative impact of higher-order corrections is significantly increased.

\newpage
\begin{figure}[t]
  \centering
  \includegraphics[width=0.45\textwidth]{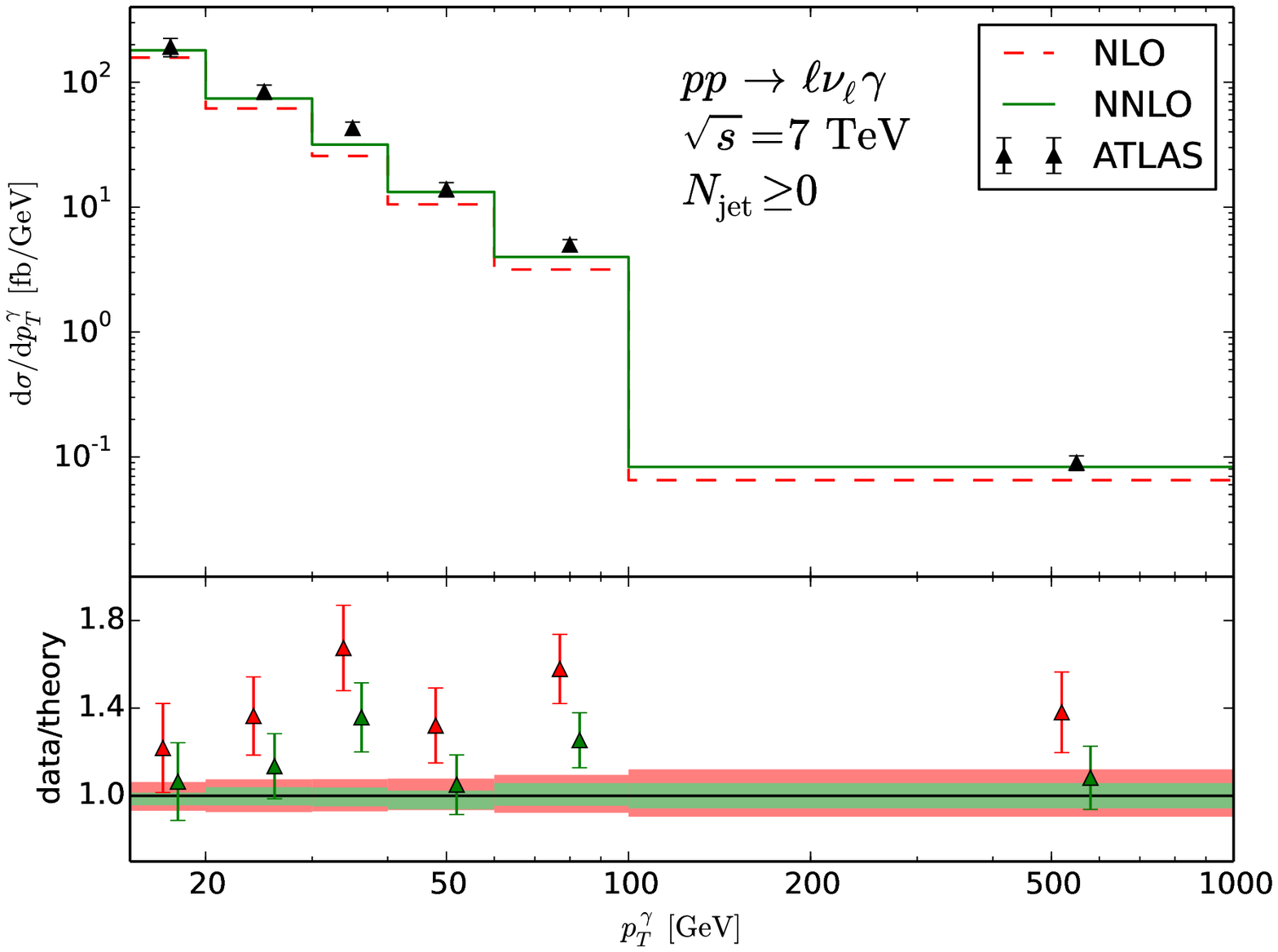}
  \hfill
  \includegraphics[width=0.45\textwidth]{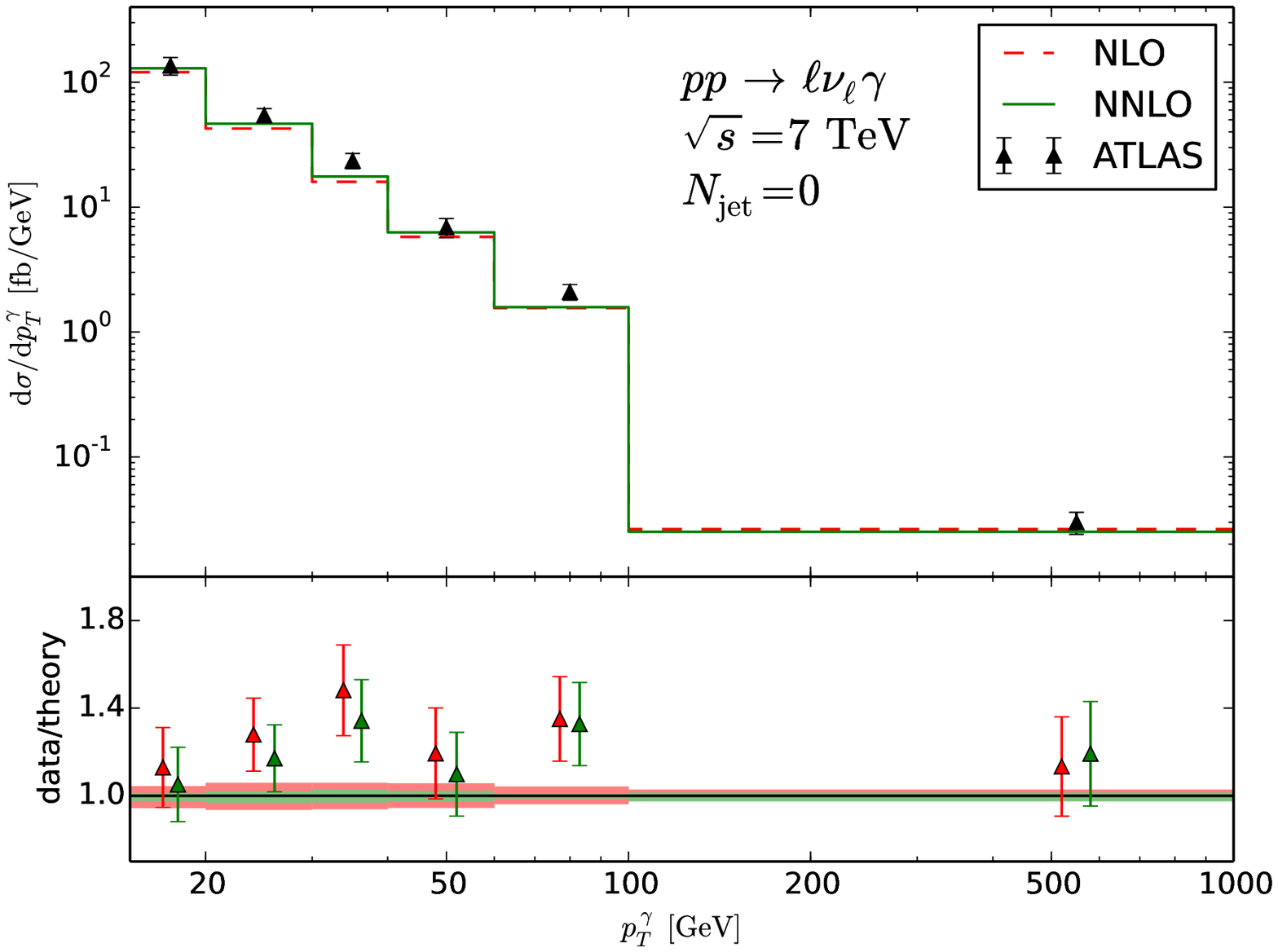}\\[2ex]
  \includegraphics[width=0.45\textwidth]{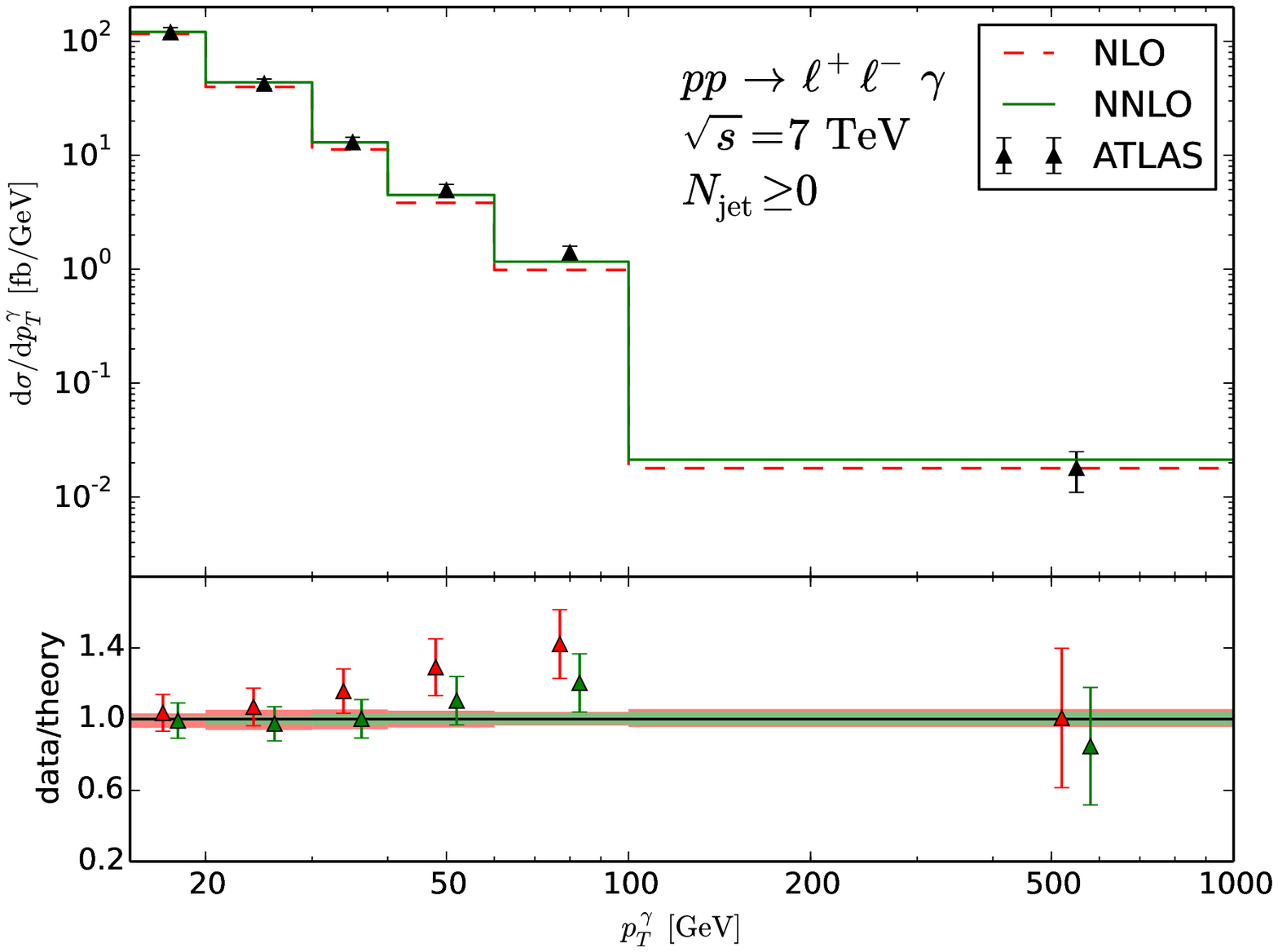}
  \hfill
  \includegraphics[width=0.45\textwidth]{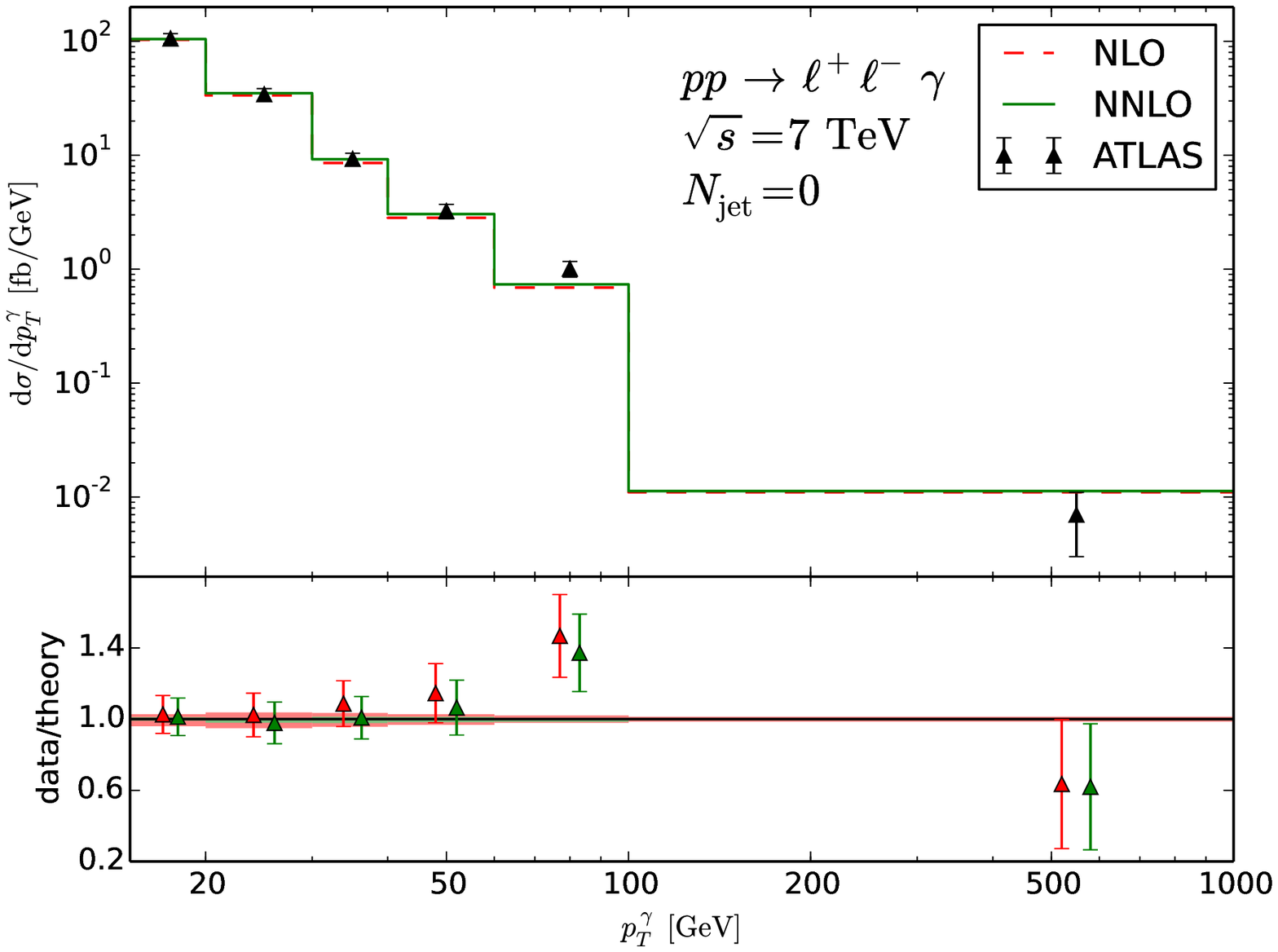}\\[-1ex]
  \caption{Photon transverse momentum distribution for the processes $pp(\to W\gamma)\to \ell\nu\gamma$ (upper plots) and $pp(\to Z\gamma)\to \ell^+\ell^-\gamma$ (lower plots) in the inclusive (left) and exclusive case (right) at NLO (red, dashed) 
and NNLO (green, solid) compared to ATLAS data. In the upper panel, only experimental uncertainties are shown. 
The lower panel shows the data/theory ratio for both theory preditions, 
and the bands indicate theoretical uncertainty estimates from scale variations.}
  \label{fig:atlas_7_pT}
\end{figure}
Beyond the cross section in the fiducial region, ATLAS has also provided the measured cross sections differential in the photon transverse momentum. 
A comparison of the resulting distributions with our theoretical NLO and NNLO predictions is displayed in \reffi{fig:atlas_7_pT} for the processes $pp(\to W\gamma)\to \ell\nu\gamma$ (upper plots) and $pp(\to Z\gamma)\to \ell^+\ell^-\gamma$ (lower plots), both for the inclusive (left plots) and the exclusive (right plots) case. 
In general, the inclusion of NNLO corrections significantly improves the 
agreement between data and theory. The improvement is particularly important in the inclusive $W\gamma$ case, and less pronounced for $Z\gamma$ and for the exclusive predictions, where the overall size of NNLO corrections is significantly smaller.\\

When switching to a harder cut of $40\;\GeV$ on $\pT^\gamma$, \refta{tab:results_VA} shows significantly increased corrections of $+242\%$ and $+79\%$ at NLO and of $+26\%$ and $+19\%$ at NNLO for the processes $pp(\to W\gamma)\to \ell\nu\gamma$ and $pp(\to Z\gamma)\to \ell^+\ell^-\gamma$, respectively.

 \begin{figure}[t]
  \centering
  \includegraphics[width=0.45\textwidth]{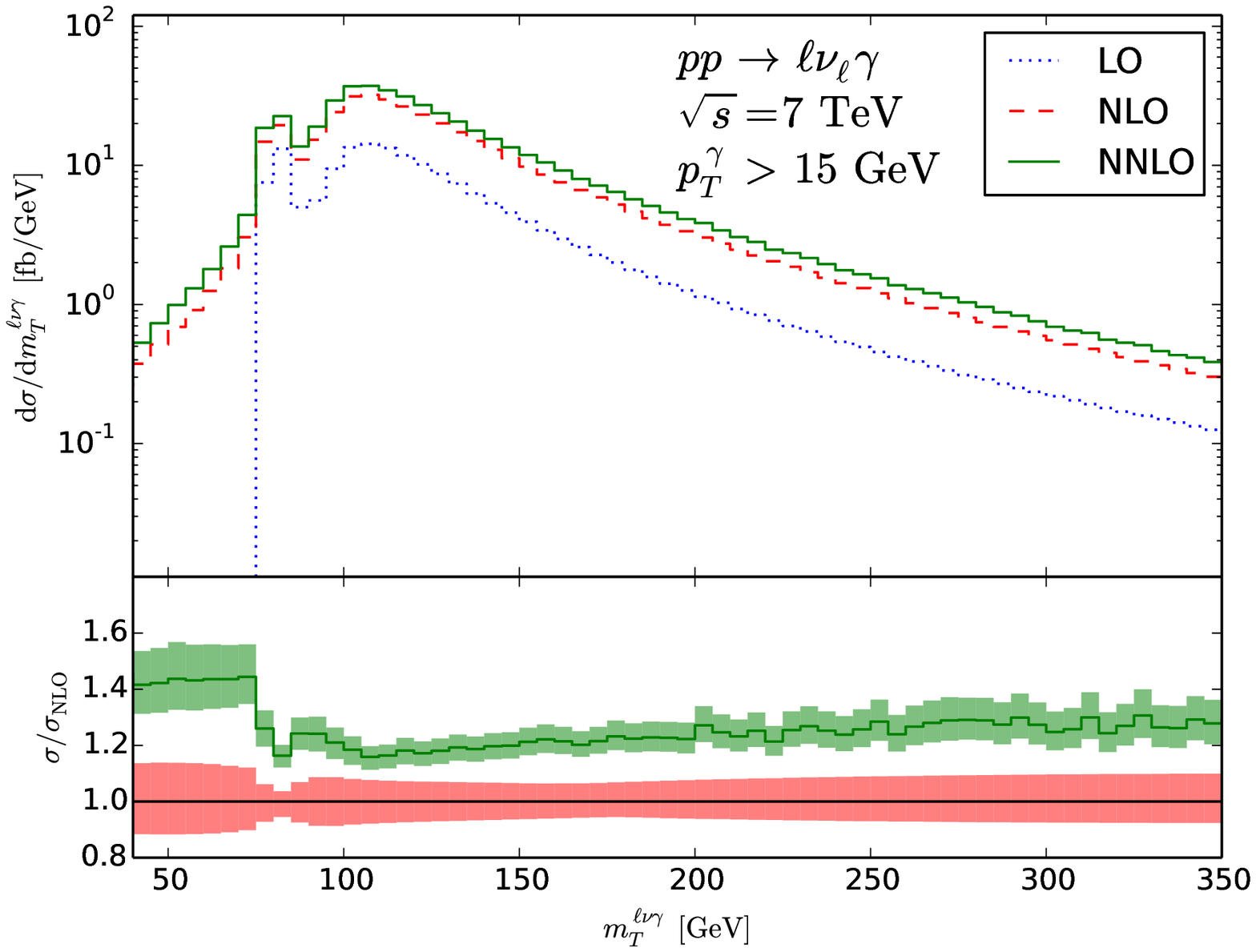}
  \hfill
  \includegraphics[width=0.45\textwidth]{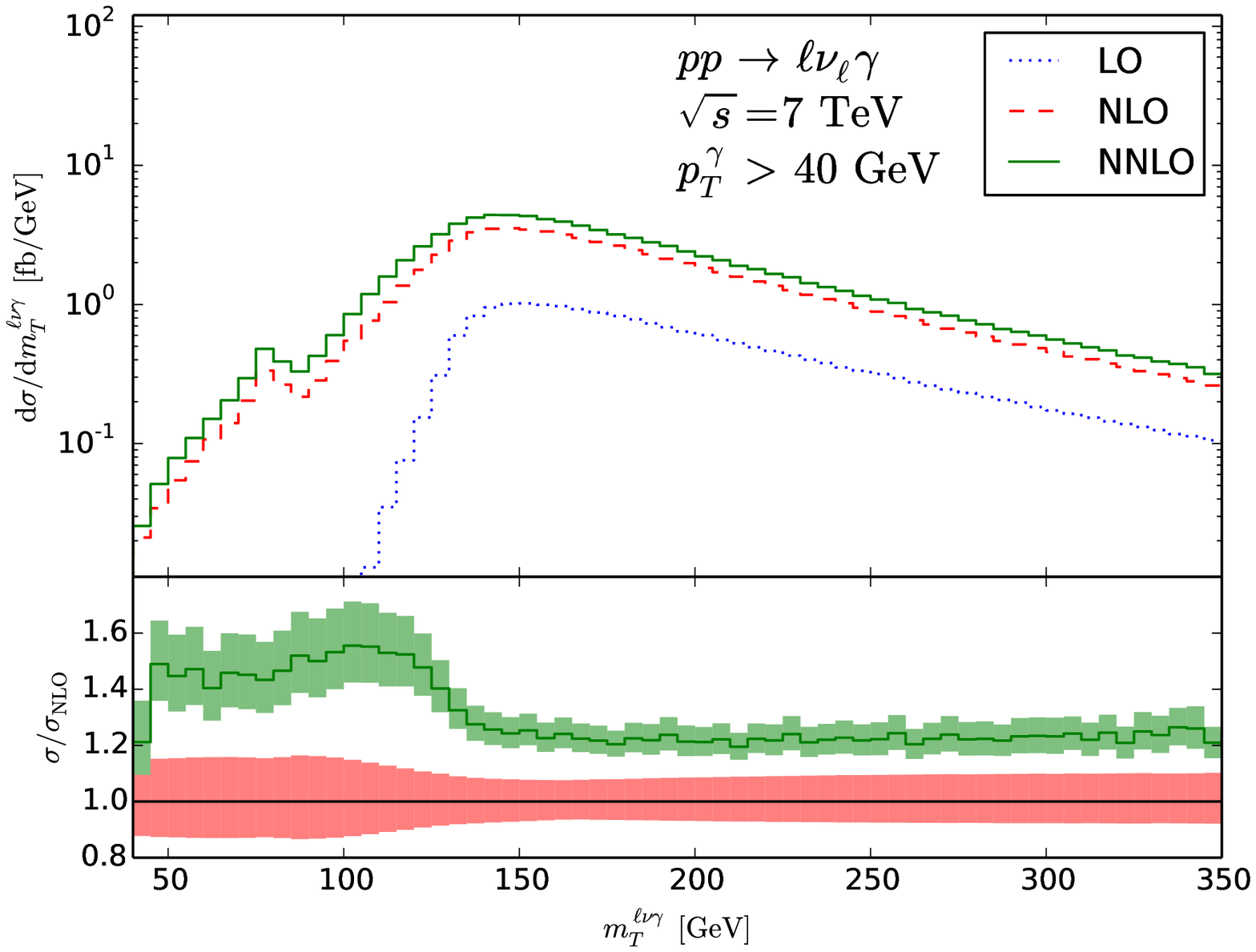}\\[2ex]
  \includegraphics[width=0.45\textwidth]{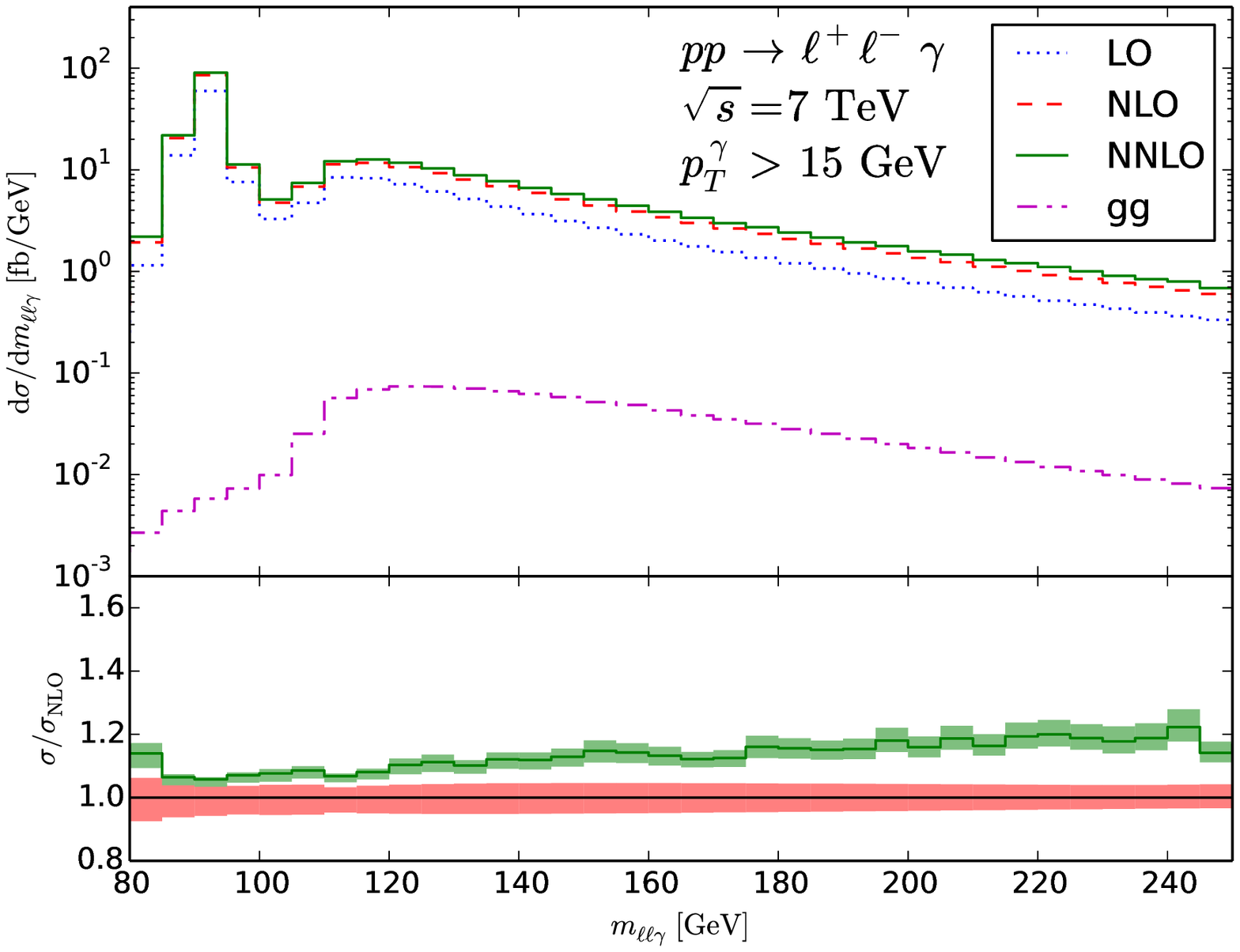}
  \hfill
  \includegraphics[width=0.45\textwidth]{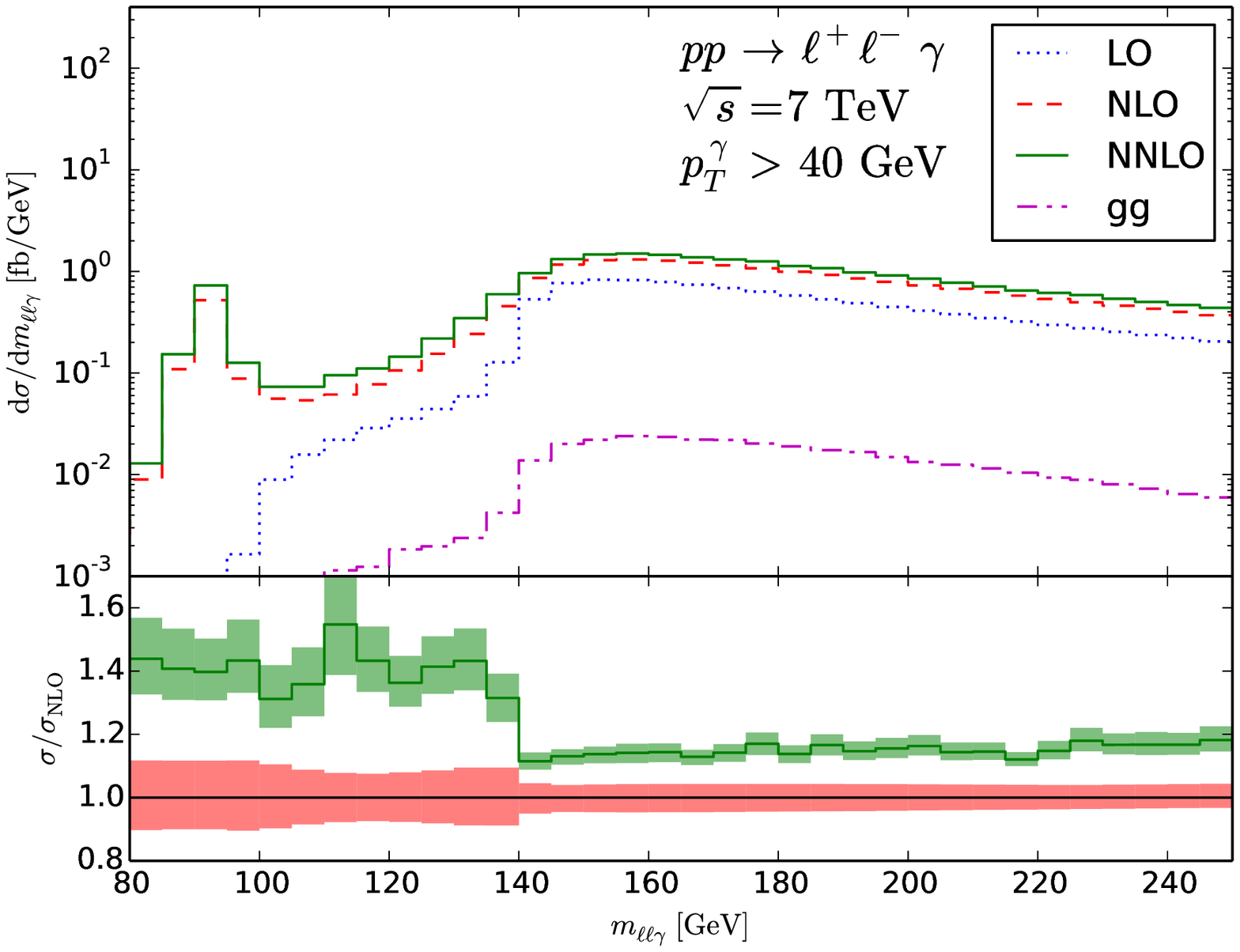}\\[-1ex]
  \caption{Transverse-mass distribution of the $\ell\nu_\ell\gamma$ system (upper plots) and invariant mass distribution of the $\ell^+\ell^-\gamma$ (lower plots) at LO (blue, dotted), 
NLO (red, dashed) and NNLO (green, solid) for $\pT^\gamma>15\;\GeV$ (left) andi 
$\pT^\gamma>40\;\GeV$ (right), in the inclusive case ($N_{\rm jet}\geq 0$). 
For $Z\gamma$ production, the loop-induced gluon fusion contribution is also shown (pink, dash-dotted). 
The lower panel shows the NNLO/NLO ratio, 
and the bands indicate theoretical uncertainty estimates from scale variations.
}
  \label{fig:7_mT_m_theory}
\end{figure}

In the $W\gamma$ case, this increased relative impact of higher-order corrections in case of a harder $\pT^\gamma$ cut can be well understood 
by studying distributions in the transverse-mass of the $\ell\nu_\ell\gamma$ system,
\begin{align}
\left(\mT^{\ell\nu\gamma}\right)^2 = \left(\sqrt{m_{\ell\gamma}^2+\left|\vec{p}_T^{\,\gamma}+\vec{p}_T^{\,\ell}\right|^2}+E_T^{\mathrm{miss}}\right)^2-\left|\vec{p}_T^{\,\gamma}+\vec{p}_T^{\,\ell}+\vec{E}_T^{\textrm{miss}}\right|^2,
\end{align}
with soft and hard $\pT^\gamma$ cuts in more detail. The corresponding plots are shown in \reffi{fig:7_mT_m_theory} (upper plots). 
For $\pT^\gamma>15\;\GeV$ (left plot), in Born kinematics the transverse mass has a lower bound of $\mT^{\ell\nu\gamma}\gtrsim 75\;\GeV$, 
i.e.\ below the $W\to\ell\nu_\ell\gamma$ peak. When the cut is increased to $\pT^\gamma>40\;\GeV$ (right plot),
this lower bound increases to $\mT^{\ell\nu\gamma}\gtrsim 100\;\GeV$,
and the $W\to\ell\nu_\ell\gamma$ peak is only populated by 
real emissions starting from the NLO. This leads to large corrections in a region where the cross section is sizeable, and thus explains the large effect on the fiducial cross section.

In the $Z\gamma$ case, an analogous reason for the increased size of corrections with a harder $\pT^\gamma$ cut can be found by studying the invariant mass distribution of the $\ell^+\ell^-\gamma$ system, which is also depicted in \reffi{fig:7_mT_m_theory} (lower plots).
For $\pT^\gamma>15\;\GeV$ (left plot), a lower bound of $m_{\ell^+\ell^-\gamma} \gtrsim 66\;\GeV$ exists in Born kinematics, i.e.\ the $Z\to \ell^+\ell^-\gamma$ peak is populated already at LO, and the region below the cut does not significantly affect the fiducial cross section. 
When the cut is increased to $\pT^\gamma>40\;\GeV$ (right plot), 
the applied cuts produce a lower bound of $m_{\ell^+\ell^-\gamma}\gtrsim 97\;\GeV$ in LO kinematics, and the $Z\to \ell^+\ell^-\gamma$ peak is not populated at all at LO.
The region below the boundary contributes sizably to the cross section, but in this region the NLO computation provides actually the leading non-vanishing prediction.
Hence the NNLO predictions effectively correspond to the first perturbative correction,
with a comparably large $K$ factor of about $1.4$.

The loop-induced gluon fusion process, also shown in \reffi{fig:7_mT_m_theory}, turns out to be small: it amounts only to around $6 (9)\%$ of the full $\mathcal{O}\left(\as^2\right)$ 
correction and, correspondingly, to less than $1(2)\%$ of the total 
fiducial cross section in case of the soft and the hard $\pT^\gamma$ cut, respectively.\\

The predicted fiducial cross sections for $pp\to \nu\bar\nu\gamma$ in the ATLAS setup 
at $7\;\TeV$~\cite{Aad:2013izg} are presented in \refta{tab:results_VA}, 
summed over three neutrino channels, and show relative corrections of $+57\%$ ($+12\%$) 
at NLO and $+12\%$ ($-2\%$) at NNLO in the inclusive (exclusive) case.
The inclusive NNLO prediction is in good agreement with the cross section measured by ATLAS. In the exclusive case, $N_{\mathrm{jet}}= 0$, the NNLO corrections are very 
small, with most likely underestimated scale uncertainties at the $1\%$ level,
and we observe quite a significant discrepancy 
with respect to the ATLAS measurement.
This can be understood by hadronization corrections, which are stated to be small for
all the other discussed processes, but lead
to sizeable effects in $\nu\bar\nu\gamma$, particularly for $N_{\rm jet}=0$.
Here, the $Z\to\nu\bar\nu$ decay implies that the final state can be identified
only through the photon and the additional radiation.
The comparison of our NLO result with that quoted
in Table VII of \citere{Aad:2013izg}, which is corrected for hadronization effects,
indeed shows that in this case an ${\cal O}(30\%)$ correction must be applied to the parton level theoretical prediction, thus reconciling it with the experimental result.

\section{Summary and discussion}
\label{sec:summary}

In these proceedings we have reported on a complete and fully differential computation of QCD radiative corrections to $W\gamma$ and $Z\gamma$ production at hadron colliders. More precisely, we have considered the processes $pp\to \ell^+\ell^-\gamma$, $pp\to \nu_\ell\overline{\nu}_\ell\gamma$ and $pp\to \ell\nu_\ell \gamma$, where, in the first case, the lepton pair $\ell^+\ell^-$ is produced either by a $Z$ boson or a virtual photon. The diagrams in which the photon is radiated off the final-state charged leptons were consistently included.
We have presented quantitative predictions for fiducial cross sections and for various kinematical distributions for $pp$ collisions at $\sqrt{s}=7\;\TeV$. 
The impact of QCD radiative corrections strongly depends on the applied cuts. In the case of $Z\gamma$, the impact of NNLO corrections is generally moderate, ranging from 8\% to 17\%. We have also shown that the loop induced gluon fusion contribution is generally small, and it accounts for less than 10\% of the full ${\cal O}(\as^2)$ correction.
In the case of $W\gamma$ production the NNLO effects are more important, and range from 19\% to 26\%. The larger impact of QCD radiative effects in the case of $W\gamma$ production is a well known consequence of a radiation zero~\cite{Mikaelian:1979nr} existing in the $W\gamma$ amplitude at Born level. This effect produces a suppression of the LO distribution in the rapidity difference between the charged lepton and the photon, and NLO and NNLO corrections are thus quite significant.
As expected, the impact of QCD radiative effects is strongly reduced when a jet veto is applied ($N_{\rm jet}=0$), being smaller than $3\%$ in the case of $Z\gamma$, and about $7\%$ in the case of $W\gamma$.

The uncertainties from missing higher-order contributions were estimated through scale variations, and turn out to be of the order of $\pm 4\%$ ($pp\to \ell\nu_\ell \gamma$), $\pm (1-2)\%$ ($pp\to \ell^+\ell^-\gamma$), and $\pm (2-3)\%$ ($pp\to \nu_\ell\overline{\nu}_\ell\gamma$) in the inclusive case (see \refta{tab:results_VA}).
Whereas the NNLO--NLO difference clearly exceeds the NLO scale band, we believe that the NNLO scale uncertainties obtained in the case $N_{\rm jet}\geq 0$ should provide the correct order of magnitude of the true uncertainty, as it is the first order at which all partonic channels are accounted for.
For $N_{\rm jet}= 0$, a more conservative approach has to be adopted to obtain a realistic estimate of the perturbative uncertainty.

The quantitative predictions we have presented for $\sqrt{s}=7\;\TeV$ were obtained by using the same cuts
adopted by the ATLAS collaboration in their measurement of the $W\gamma$ and $Z\gamma$ cross sections~\cite{Aad:2013izg}.
We compared to ATLAS data, both for the fiducial cross sections and for some kinematical distributions, and the agreement between data and theory is in general improved at NNLO, in particular the former $\approx2\sigma$ excess in $W\gamma$ compared to NLO is reduced well below $1\sigma$ (the remaining discrepancy to our prediction for $pp\to \nu_\ell\overline{\nu}_\ell\gamma$ with a jet veto is understood). 

To achieve reliable predictivity in the high-$\pT^\gamma$ region, a combination of our results with EW corrections~\cite{Denner:2014bna,Denner:2015fca} is required, which is, however, left for future work.

\section*{Acknowledgments}
This research was supported in part by the Swiss National Science Foundation (SNF) 
under contracts CRSII2-141847, 200021-144352, 200021-156585 and by the Research Executive Agency (REA) of the 
European Union under the Grant Agreement number PITN-GA-2012-316704 (\mbox{\it Higgstools}).

\end{document}